\documentclass[achemso, graphicx, reprint]{revtex4-2}

\usepackage{graphicx}
\usepackage{dcolumn}
\usepackage{bm}

\usepackage[utf8]{inputenc}
\usepackage[T1]{fontenc}
\usepackage{mathptmx}
\usepackage{etoolbox}
\usepackage{color}
\usepackage{amssymb}
\usepackage{amsmath}

\makeatletter
\def\@email#1#2{%
 \endgroup
 \patchcmd{\titleblock@produce}
  {\frontmatter@RRAPformat}
  {\frontmatter@RRAPformat{\produce@RRAP{*#1\href{mailto:#2}{#2}}}\frontmatter@RRAPformat}
  {}{}
}%
\makeatother

\setcitestyle{super}

\begin{document}


\title{Deterministic generation of single B centers in hBN by one-to-one conversion from UV centers} 


\author{Andr\'es N\'u\~nez Marcos, Christophe Arnold, Julien Barjon, St\'ephanie Buil, Jean-Pierre Hermier, Aymeric Delteil}


\affiliation{Universit\'e Paris-Saclay, UVSQ, CNRS, GEMaC, 78000, Versailles, France.\\ 
{\color{white}---------------------------------------} aymeric.delteil@usvq.fr{\color{white}---------------------------------------} }



\begin{abstract}
\vspace{0.2cm}
\section*{Abstract}
Among the variety of quantum emitters in hexagonal boron nitride (hBN), blue-emitting color centers, or B centers, have garnered a particular interest owing to their excellent quantum optical properties. Moreover, the fact that they can be locally activated by an electron beam makes them suitable for top-down integration in photonic devices. However, in the absence of a real-time monitoring technique sensitive to individual emitters, the activation process is stochastic in the number of emitters, and its mechanism is under debate. Here, we implement an \textit{in-situ} cathodoluminescence monitoring setup capable of detecting individual quantum emitters in the blue and ultraviolet (UV) range. We demonstrate that the activation of individual B centers is spatially and temporally correlated with the deactivation of individual UV centers emitting at 4.1~eV, which are ubiquitous in hBN. We then make use of the ability to detect individual B center activation events to demonstrate the controlled creation of an array with only one emitter per irradiation site. Additionally, we demonstrate a symmetric technique for heralded selective deactivation of individual emitters. Our results provide insights into the microscopic structure and activation mechanism of B centers, as well as versatile techniques for their deterministic integration.\\

Keywords: hexagonal boron nitride; color centers; cathodoluminescence; 2D materials; single-photon emission; quantum photonics

\end{abstract}

\pacs{}
\maketitle 


Optically active point defects in crystals, commonly referred to as color centers, are of great significance in quantum information science due to their ability to act as single-photon emitters (SPEs) and their compatibility with nanostructures and devices.\cite{Aharonovich2016} A well-known example is the nitrogen-vacancy (NV) center in diamond, which forms the foundation of many emerging quantum network architectures.\cite{Hanson2013} More recently, the discovery of color centers in van der Waals materials has opened new avenues for integration and applications in quantum technologies,\cite{Moody2021} with hexagonal boron nitride (hBN) being a particularly promising host. This wide-bandgap semiconductor can indeed integrate high quality SPEs.\cite{Tran16,Martinez16} A specific class of hBN emitters is distinguished by its remarkable quantum optical properties. The so-called B centers reproducibly emit at 436~nm and exhibit photon indistinguishability at low temperature.\cite{Fournier23PRA, Gerard25} They are not found natively but can be locally activated by a focused electron beam,\cite{Shevitski19, Fournier21, Gale22} which is an asset for the controlled integration in photonic devices.\cite{Gerard23, Spencer23} However, the activation mechanism is still debated and yields an unpredictable number of SPEs, due to the lack of a technique for heralding the successful creation of single emitters.

The microscopic structure of the B centers is still under debate. Some important clues have been however identified. In particular, the B center activation is efficient only in crystals exhibiting the optical signature of UV centers.\cite{Gale22} These UV color centers emitting at 4.1~eV are commonly found in carbon-rich hBN\cite{ Taniguchi07, Silly07, Museur08, Bourrelier16, Schue17, Onodera19} and are attributed to in-plane carbon dimers (C$_\mathrm{B}$C$_\mathrm{N}$).\cite{Mackoit19, Li22,Plo25} Cathodoluminescence (CL) has been shown to allow for real-time monitoring of the emitter population dynamics under electron irradiation, allowing to study the kinetics of B center creation\cite{Roux22} and to observe the concurrence of a decrease in the population of UV centers,\cite{Nedic24} albeit with a different timescale. These studies have been performed on ensembles and at room temperature, therefore the possible relation between UV and B centers remains unclear. Moreover, the activation and deactivation of presumably individual B centers has been observed,\cite{Shevitski19,Hou25} but without insights about the mechanism of such dynamics and the relation to individual UV centers. Concerning the microscopic structure of B centers, several candidates have been proposed, such as a split nitrogen interstitial,\cite{Zhigulin23, Ganyecz24} a carbon tetramer,\cite{Maciaszek24, Gale25} and a vertical carbon dimer.\cite{Zhigulin23} The latter hypothesis seems to have been confirmed very recently by a bundle of consistent clues obtained with combined STEM/HAADF and CL measurements.\cite{Hou25}
 
 \begin{figure*}
 \includegraphics[width=\linewidth]{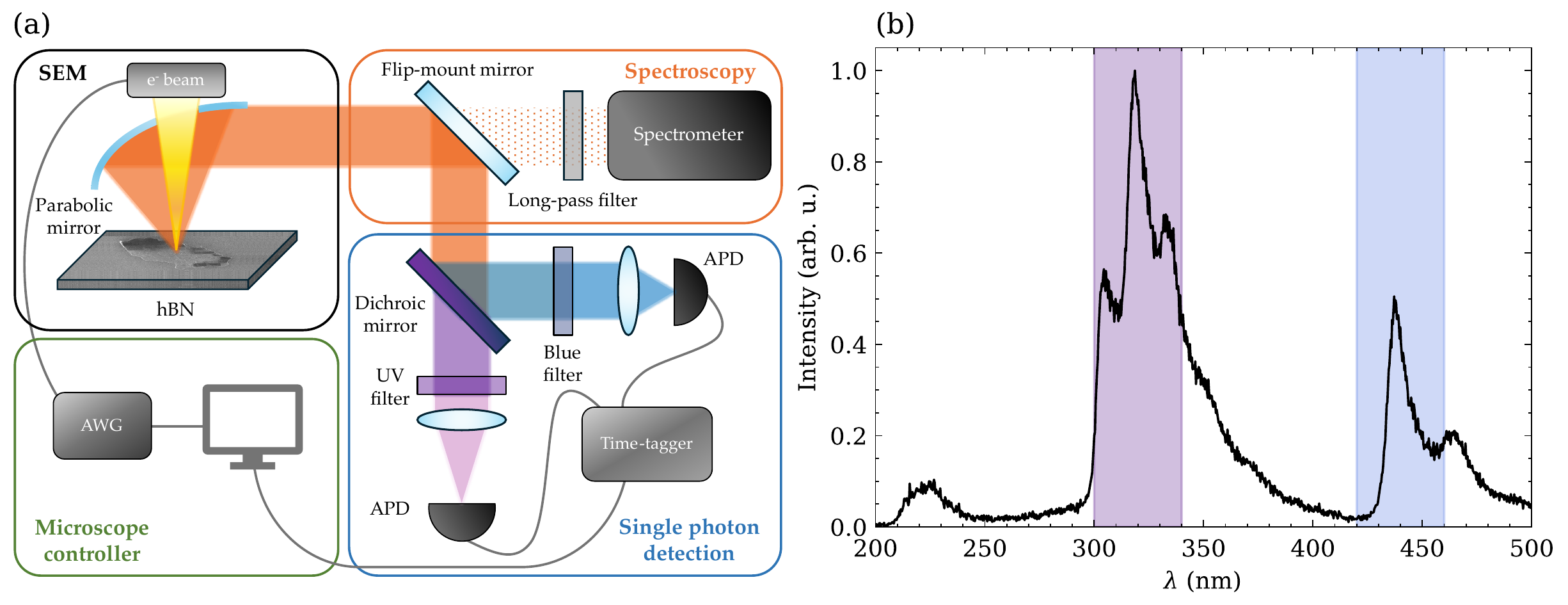}
 \caption{\label{Setup} (a) Scheme of the experimental setup. AWG: arbitrary waveform generator. (b) Black line: CL spectrum of a hBN sample containing UV and blue emitters. Violet shading: UV filter passband. Blue shading: Blue filter passband. The visible-range part of the spectrum was taken with a UV filter to suppress second-order refraction of the exciton emission.}%
 \end{figure*}
 
Here, we investigate the activation and deactivation of individual B and UV centers upon irradiation by an electron beam, through a specifically designed CL detection setup based on avalanche photodiodes (APDs), which makes the \textit{in-situ} monitoring sensitive to the single emitter level in both wavelength ranges. We show that individual creation events of B centers are accompanied by a simultaneous abrupt deactivation of a UV center, correlated both in time and space. This suggests a one-to-one transformation of UV to B centers. The conversion is partially reversible, as the opposite process is also observed. At long times, B centers stabilize into a steady state population, as demonstrated in prior work.\cite{Roux22, Nedic24} In a second step, we make use of the \textit{in-situ} monitoring signal as a marker that heralds successful creation of individual emitters. We demonstrate a small array of emitters with strongly sub-Poissonian statistics of emitter numbers, as confirmed with the measurement of their second-order correlation function. We then show that accidental double creations can be corrected for using heralded photobleaching, further reducing the variations in emitter numbers. In a conclusive discussion, we tentatively provide a mechanism for supporting the observed UV-blue conversion, consistent with the vertical carbon dimer hypothesis for the microscopic structure of B centers.

\section*{Interconversion of UV and B centers}
We investigate hBN exfoliated from high-pressure, high-temperature grown crystals natively incorporating carbon impurities (see Methods for sample preparation). The experimental setup which allows for \textit{in-situ} monitoring of the activation and deactivation of individual color centers in the UV and blue spectral range is presented in Figure~\ref{Setup}a. The sample is placed inside a scanning electron microscope (SEM), where a holed parabolic mirror is positioned above the sample to collect the emitted photons outside the microscope. A spectrometer is placed at the output port of the SEM. Figure~\ref{Setup}b shows the CL spectrum of a hBN crystal under electron irradiation. Three recognizable features can be observed. The exciton emission is visible at about 215~nm; the three equally spaced peaks in the 300-350~nm range are the signature of the UV emitter C$_\mathrm{B}$C$_\mathrm{N}$ with a zero-phonon line at 305~nm and its phonon replica; the peaks at 440~nm and 465~nm originate respectively from the B center zero-phonon line and optical phonon replica.\cite{Taniguchi07, Vuong16, Plo25} Using a flip-mount mirror, we can channel the emission to a single-photon detection system. The emitted light is then separated into two frequency bands, UV and blue regions respectively, using a dichroic mirror and bandpass filters.  The bandwidth of both detection channels is depicted by the shaded areas on Fig.~\ref{Setup}b. Both signals are simultaneously detected by APDs. This setup allows to be sensitive to the CL signal associated with individual UV and B centers.

 \begin{figure*}
 \includegraphics[width=\linewidth]{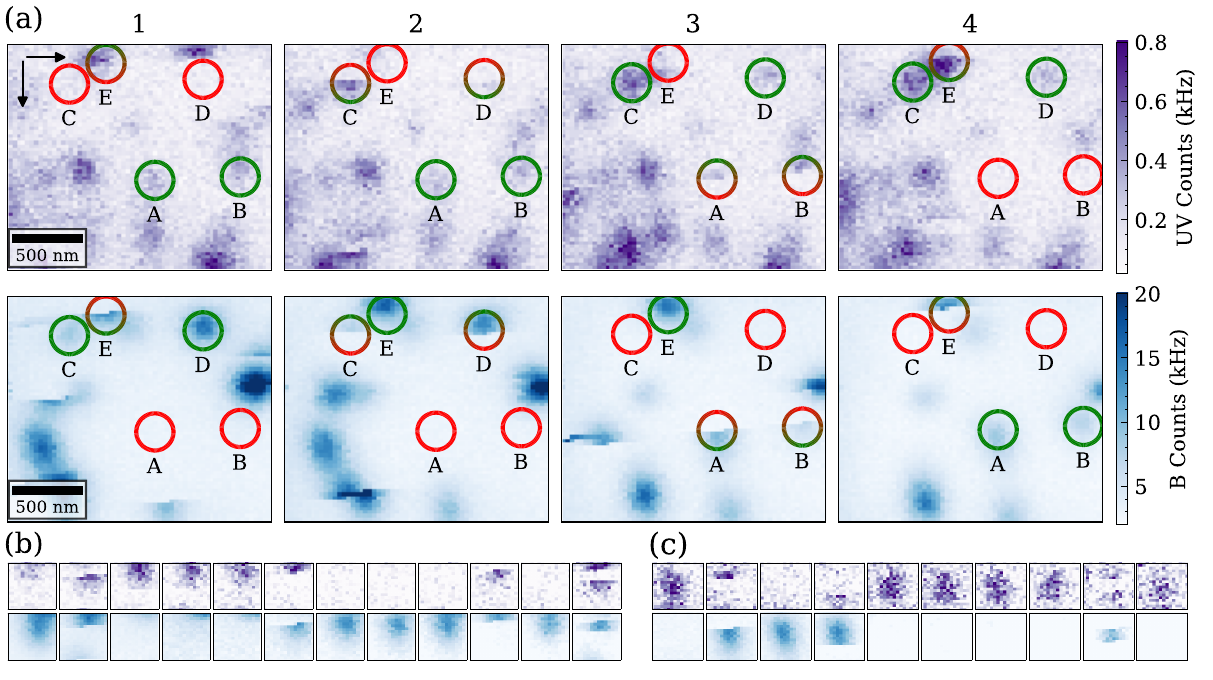}%
 \caption{\label{CL_map} (a) Series of CL bandpass images of color centers in a hBN flake of thickness 145~nm with a current intensity of $I = 0.024$~nA. The top row shows the signal measured in the UV detection channel and the bottom row displays the signal measured in the blue detection channel. The scale bar indicates 500~nm. The circles labeled A to E indicate the state of five emitters: a red circle indicates the color center is off, a green circle indicates the emitter is on, and a two-colored circle indicates that an activation or deactivation is taking place. The states of the blue and UV emitters occupying these five positions are anticorrelated. (b) and (c) Consecutive CL scans of two fixed emitters in the UV (top) and blue (bottom) spectral range, showing switching between the two states.}%
 \end{figure*}

We first select a hBN crystal of thickness 145~nm, where the density of UV centers is low enough to spatially resolve individual emitters. Figure~\ref{CL_map}a shows a series of four consecutive CL scans of the same region of space. The map size is $1.6$~$\mu$m $\times 1.9$~$\mu$m, corresponding to $64 \times 64$~pixels. The scanning time is about ten minutes, and the scan direction is from left to right, then from top to bottom. Both UV and blue channels are recorded in parallel, simultaneously. The top (bottom) panels plots the UV (blue) band signal. We refer to these successive maps as scan 1 to 4. In both spectral ranges, various point emitters can be observed. In addition, some emitters of both species either appear or disappear during scans. To ease the discussion, we select five representative sites, depicted by circles in both spectral regions and labeled by letters A to E. The circle colors indicate that the emitters are active (green circles), inactive (red circles) or switching (two-colored circles). In the UV maps, sites A and B indicate UV emitters that are active during scans 1 and 2 (upper panel), while no blue emitter is observed at these same sites (lower panel). In scan 3, the UV emitters located sites A and B are deactivated, which shows up as interrupted spots with CL signal only in the upper part. Simultaneously, signal appears in the blue range at the same locations. The switching appears instantaneous at the timescale of the integration time per pixel (100 ms), as the blue signal appears at the same pixels where the UV signal disappears. In scan 4, no UV emitter is visible in sites A and B, while blue emitters now appear as full spots. Sites C and D exhibit the inverse process along the four scans, with blue emitters initially present in scan 1, switching to UV during scan 2, and staying visible as UV emitters in the two following scans. Finally, site E undergoes a bidirectional process, with an initial switching from UV to blue during scan 1, staying in the blue state during scans 2 and 3, and switching back to UV during scan 4. Other similar events can be observed on these CL scans, as well as all other scans we performed. We illustrate this further by showing consecutive scans of two emitters on Figure~\ref{CL_map}b and~c. Both emitters exhibit multiple switching events between the UV and the blue state while the electron beam is passing over them. Additional examples of consecutive scans are shown in the Supporting Information section~S1. 
We note that the irradiation parameters are chosen so that only a few emitters switch at every scan, so to make the observation of the interconversion more evident. Since it is a stochastic process, some UV emitters are expected to persist during many frames. However, it cannot be excluded that some of the persistent spots observed on the UV maps originate from other defect species. CL spectra suggest however that the UV emission is largely dominated by contribution from the carbon dimers, and therefore that most of the UV signal originates from UV centers. On the other hand, in the blue wavelength range, regions that have not been exposed to the electron beam are initially devoid of CL signal, so that all spots are attributed to B centers (see Supporting Information section~S1). Overall, our observations can be summarized as follows.

(1) The activation of B centers is observed simultaneously with the deactivation of UV centers.

(2) The reverse process is also observed, demonstrating that the conversion is at least partially reversible.

(3) The conversion is faster than our integration time per pixel (100~ms). 

(4) The switching occurs exactly under the electron beam, as testified by truncated CL spots, with the signal appearing or disappearing at the emitter position. This can be observed as incomplete lines of bright pixels switching while the electron beam is scanning over the SPE. This implies that the process is driven locally by the electron beam, at the immediate vicinity of the UV centers. This also demonstrates that, in samples with high concentration of UV centers, the resolution of the B center creation is solely given by the SEM resolution.

(5) Essentially all occurrences where a single UV center disappears while scanning are associated with the appearance of a B center. This suggests that the conversion to B center is the main -- if not the only -- decay process of the UV center under electron irradiation. Other mechanisms were proposed, which consist in the dissociation of the C$_\mathrm{B}$C$_\mathrm{N}$ dimers into (C$_\mathrm{B}$,C$_\mathrm{N}$) donor-acceptor pairs at various neighboring positions.\cite{Bianco25} Given that such mechanisms would lead to the disappearance of UV centers without simultaneous activation of a B center, which we do not observe in our datasets, we conclude that they are likely slower or less probable, at least in our range of irradiation parameters.

The ability to observe the activation of individual emitters opens the way to the deterministic fabrication of a single one, which is what we investigate in the next section.

\section*{Heralded activation of B centers}

 \begin{figure}
 \includegraphics[width=\linewidth]{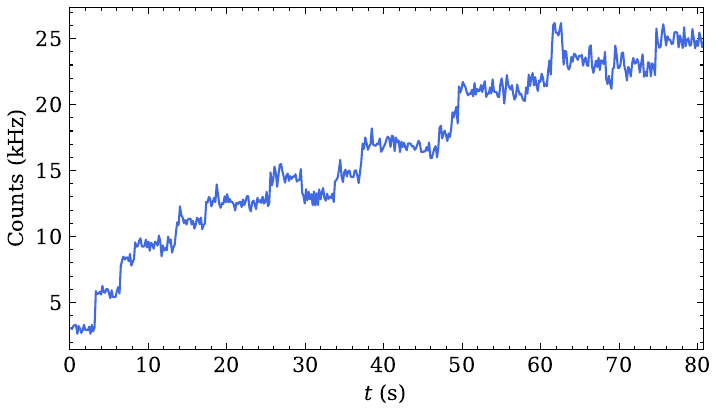}%
 \caption{\label{Timetrace} Representative CL timetrace measured by the blue detection channel during a continuous irradiation of a fixed location on a 130~nm flake with $I = 0.021$~nA.}%
 \end{figure}
  
 \begin{figure*}
 \includegraphics[width=\linewidth]{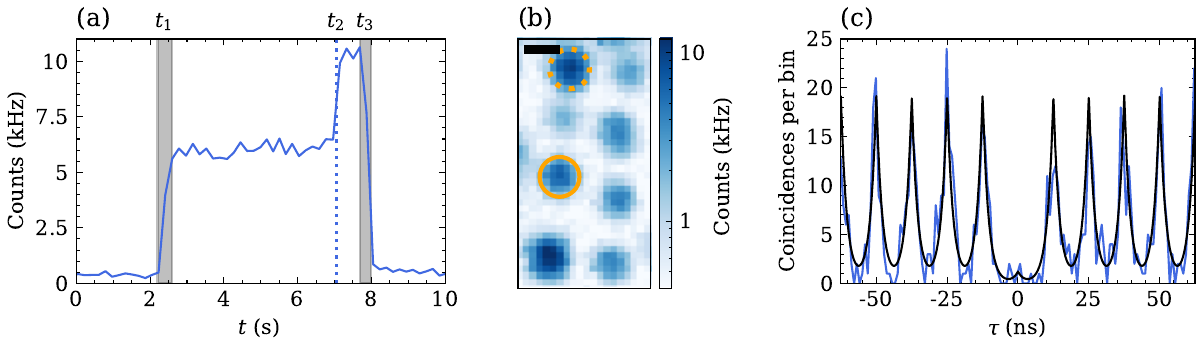}%
 \caption{\label{Array} (a) CL intensity trace of the blue detection channel: at $t_{1}$ the $e^{-}$ beam is turned on with a current of 0.021~nA; at $t_{2}$ a single B center is generated and at $t_{3}$ the $e^{-}$ beam is turned off, exposing the sample to a dose of approximately $4.5 \times 10^{8}$ electrons. (b) PL map of a $4 \times 2$ array of B centers activated in a 50~nm thick flake, with a distance of 1~$\mu$m between irradiation sites. The scale bar indicates 500~nm. The orange circle indicates the individual B center characterized in Fig.~\ref{Array}c and the dotted orange circle denotes the irradiation site characterized in Fig.~\ref{Bleaching}.  (c) Blue line: $g^{(2)}(\tau)$ measured for the B center circled in orange in the PL map. Black line: fit to the data, providing $g^{(2)}(0) = 0.09 \pm 0.03$.}%
 \end{figure*}

We now select a flake with a higher density of UV centers, such that conversion can occur at any location. Such flakes exhibit a large UV signal, similar as the spectrum shown Fig.~\ref{Setup}b. We continuously irradiate the crystal at a fixed position, while recording the signal of B centers with the APD. Figure~\ref{Timetrace} shows a representative example of the intensity trace measured during 80~s. Discrete increases of the signal can be observed, showing that individual creation events can be unambiguously identified. The signal associated with the creation of individual emitters is about $10^3$ to $10^4$ counts per second. A few downward steps are also visible, associated with deactivations as observed in the previous section. However, the overall trend is a global increase of the number of emitters, consistently with prior work on large ensembles.\cite{Roux22, Nedic24} This suggests that the blue emitters are stabilized under continuous irradiation. Each upward step constitutes a signal heralding the success of individual B center activation.

We illustrate the potential use of the individual activation monitoring by realizing an array of single emitters as a proof of principle. We irradiate at eight chosen locations while monitoring the signal, and we interrupt the irradiation after the first upward step is observed. An example of real-time signal is shown Figure~\ref{Array}a. The protocol is the following:

(1) We switch on the electron beam at $t_1$, from where a background luminescence signal of about 5 kHz is observed.

(2) At time $t_2$, an upward step is observed.

(3) Shortly after, the electron beam is switched off manually, at time $t_3$, using the built-in blanking system of the SEM.

(4) We repeat this process at eight different locations in a $4 \times 2$ array, with 1~$\mu$m distance between sites. 

The sample is subsequently characterized in photoluminescence (PL) at room temperature, in a confocal microscope (see Methods). A pulsed laser emitting at 405~nm is used to excite the emitters. Figure~\ref{Array}b shows a PL confocal map of the sample. Eight luminescent spots are visible. The slight misalignment of the spot array originates from hysteresis of the piezoelectric positioner used to define the irradiation sites. The emitters are characterized in a Hanbury Brown and Twiss setup to estimate the number of emitters. The second-order correlation function $g^{(2)}(\tau)$ is measured as the histogram of the delay time between consecutive photon detection events, and fitted using a multipeak function (see Methods). A value of $g^{(2)}(0) < 0.5$ is attributed to a single emitter. An example of $g^{(2)}(0)$ is shown on Figure~\ref{Array}c, corresponding to the circled spot on Figure~\ref{Array}b. Among the eight sites, seven are single emitters, and one, indicated by the dotted circle on Figure~\ref{Array}b, has a higher $g^{(2)}(0)$. The latter corresponds to an event where two steps were observed before the electron beam was blanked. A workaround for such cases is proposed in the next section.

This proof of principle demonstrates the ability to control the number of activated color centers. Without the active feedback developed in this work using \textit{in-situ} CL monitoring, a Poisson distribution of the emitter numbers would have been obtained, yielding single centers in at most 37~\% of the irradiated spots, which is the theoretical higher bound allowed by the Poisson distribution. We note that, in the absence of \textit{in-situ} monitoring, a yield of 33~\% has been obtained using calibrated irradiation.\cite{Gale22} In the following, we demonstrate an additional step to narrow down the distribution further in case of undesired double activation, as for one of the eight sites here.

\section*{Controlled deactivation by photobleaching}

\begin{figure*}
 \includegraphics[width=\linewidth]{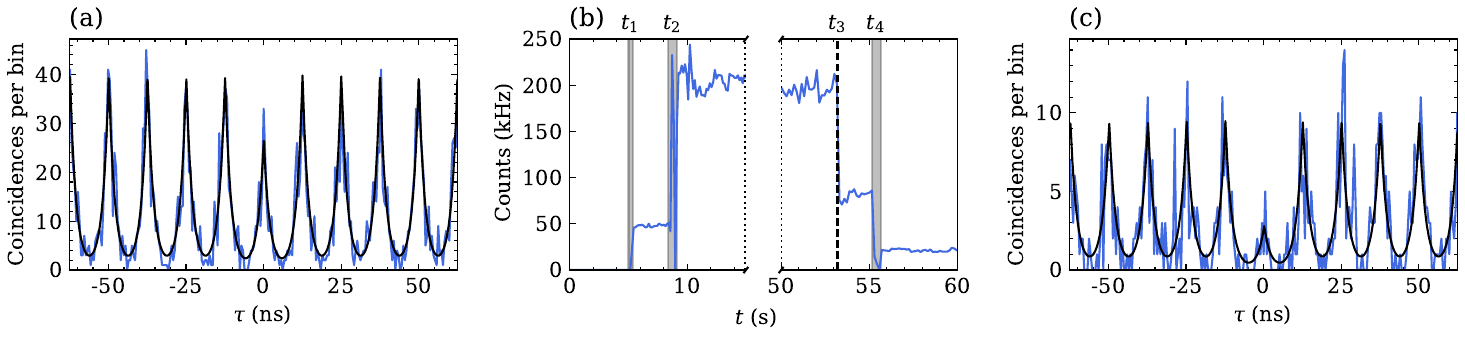}%
 \caption{\label{Bleaching} (a) Blue line: $g^{(2)}(\tau)$ measured from the brightest B-center in the PL map (dashed orange circle in Figure~\ref{Array}{b}). Black line: Fit to the data, where $g^{(2)}(0) = 0.67 \pm 0.05$ (b) PL timetrace measured during the selective photobleaching procress; $t_{1}$: the laser is turned on with a neutral density filter, yielding a 750~$\mu$W power; $t_{2}$: the absorption filter is removed, so that the excitation power increases to 7.5~mW; $t_{3}$: a B-center has been deactivated; $t_{4}$: the absorption filter is inserted back. (c) Blue line: $g^{(2)}(\tau)$ measured at the same position after the photobleaching process. Black line: fit to the data, providing $g^{(2)}(0) = 0.29 \pm 0.11$.}%
 \end{figure*}

A symmetrical approach to the activation protocol can be used for selective deactivation. B centers are very stable under laser illumination at low to moderate powers, up to about 1-2~mW, as testified by week-long measurements in PL and resonant excitation.\cite{Fournier23PRA, Gerard25} This power regime is therefore considered as ``safe'' for the emitters. However, the B centers can photobleach when exposed to a continuous-wave (cw) PL laser at high power ($\gtrsim 4$~mW) during a few tens to hundreds of seconds.\cite{Gerard24} We demonstrate how this phenomenon can be used to selectively remove an emitter from an irradiation site where two emitters are present, using cw laser drive in the high-power, ``unstable'' regime. We base our approach on real-time monitoring of the PL during laser illumination, in analogy with the activation case. To illustrate this technique, we consider the site of the array shown on Figure~\ref{Array}b (dotted orange circle) that exhibits $g^{(2)}>0.5$. Figure~\ref{Bleaching}a shows the $g^{(2)}$ of this site before selective bleaching, measured in the pulsed regime. The center peak has a value of $g^{(2)}(0) = 0.67 \pm 0.05$. We apply the selective photobleaching technique while monitoring the PL signal, shown on Figure~\ref{Bleaching}b. The process is as follows:

(1) The cw laser is switched on at time $t_1$, with a power of 0.75~mW, in the safe power regime. The count rate  is about 50~kHz.

(2) The power is abruptly raised to 7.5~mW, in the ``unstable'' regime, by removing a neutral density filter at time $t_2$. The count rate increases to 200~kHz, above saturation.

(3) At time $t_3$, a downward step is observed, heralding bleaching of an emitter. The count rate is about 80~kHz.

(4) Shortly after, we put back the neutral density filter at time $t_4$ to return in the safe power regime. The count rate is then about 20~kHz.

To verify that the remaining signal comes from a single emitter, we measure the $g^{(2)}$ function again. The result is shown on Figure~\ref{Bleaching}c. The value of  $g^{(2)}(0) = 0.29 \pm 0.11$ testifies that only one emitter is now active on the site. Further examples of controlled bleaching are shown in the Supporting Information section~S2, to demonstrate the reproducibility of the method. We note that bleaching occurs after a few tens of seconds, making it straightforward to isolate a single bleaching event. The combined monitoring of SPE activation using CL, and deactivation using PL, allows us to reach 100~\%  success for the creation of single B centers.

 \begin{figure}[b]
 \includegraphics[width=0.85\linewidth]{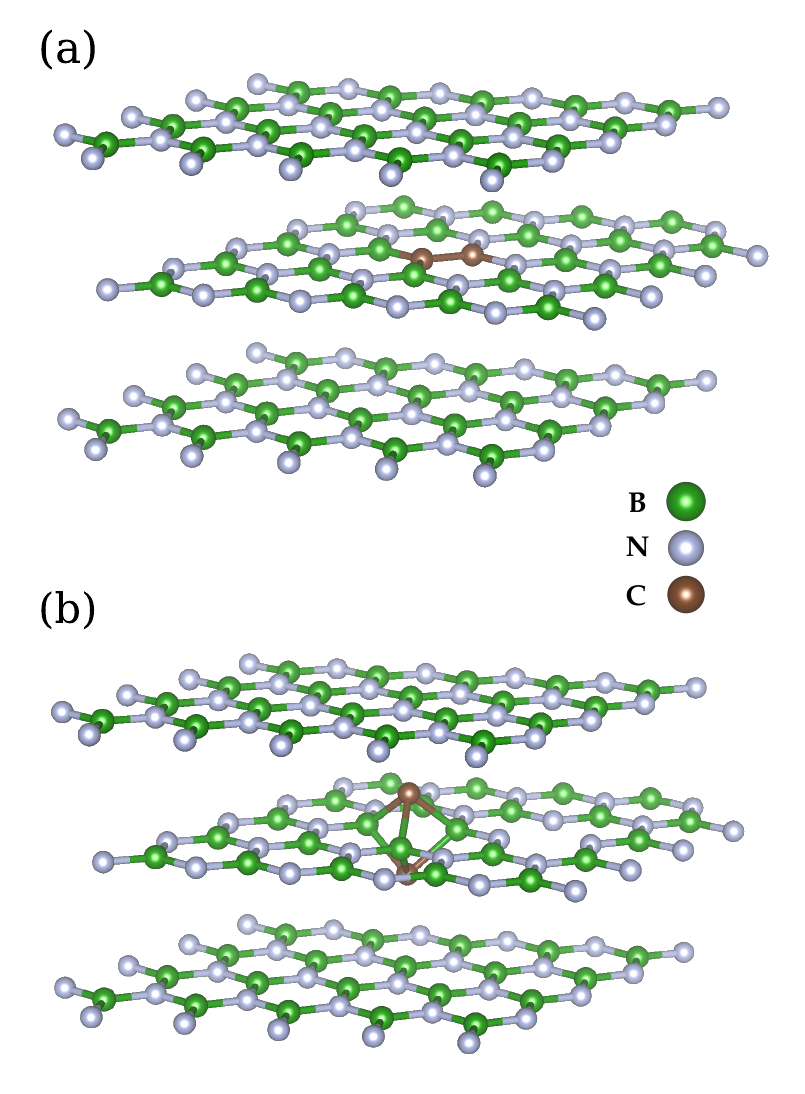}%
 \caption{\label{Atomic structures} Possible microscopic models for the studied defects. (a) In-plane carbon dimer $\mathrm{C_B C_N}$ for the UV center and (b) carbon dimer in split interstitial configuration $\mathrm{(2 C_i)}_\mathrm{N}$ for the B center, with a leftover vacancy after conversion.}%
 \end{figure}

\section*{Discussion about the conversion mechanism}

The ability to simultaneously resolve UV and B centers at the individual level has allowed us to observe electron-beam induced direct interconversion between these two species. Such events can shed additional light on the microscopic structure of both defects. Recent experimental studies identify the UV defect as a horizontal carbon dimer (C$_\mathrm{B}$C$_\mathrm{N}$)\cite{Plo25} (Fig.~\ref{Atomic structures}a) and the B center as a vertical carbon dimer in a split interstitial configuration (Fig.~\ref{Atomic structures}b), either at a boron site $\mathrm{(2C_i)}_\mathrm{B}$ or at a nitrogen site $\mathrm{(2 C_i)}_\mathrm{N}$.\cite{Hou25} These proposed atomic structures for both species have also been investigated using density functional theory (DFT) calculations,\cite{Mackoit19, Li22, Hou25} giving additional credit to these structural models. In the light of these conclusions, our results can be interpreted as the observation of electron-beam induced modification of a carbon complex, and provides additional credit to both identifications. Indeed, no other pair of candidates for either UV or B centers can be easily linked together among the plausible candidates proposed in the literature.\cite{Mackoit19, Hamdi20, Li22, Ganyecz24, Maciaszek24, Plo25, Hou25} However, a subtlety stands in the fact that the vertical dimer structure replaces only one atom of the hBN lattice, while the horizontal dimer occupies two atomic sites (Fig.~\ref{Atomic structures}). Therefore, the electron-beam activation process of the B center is not a simple 90$^\circ$ rotation of the in-plane carbon dimer, and necessarily involves an additional species.

To account for the additional atom in the B center structure as compared with the UV center, we tentatively propose the following scenario, where a vacancy is created when switching from the horizontal to the vertical configuration of the carbon doublet. 

\begin{equation}
\mathrm{C_B C_N} \rightleftharpoons \mathrm{(2 C_i)}_\mathrm{N} + V_\mathrm{B}
\end{equation}

where we have supposed that the B center is at the nitrogen site, \textit{i.e.} $\mathrm{(2 C_i)}_\mathrm{N}$, without loss of generality since the boron site counterpart can be obtained by permuting N and B. The conversion is reversible according to our observations. However, the vacancy is known to be mobile.\cite{Weston18, Alem11} Upon diffusion (which can be induced by the electron beam\cite{Alem11}), the vacancy migration prevents backwards conversion and locks the defect to the $\mathrm{(2 C_i)}_\mathrm{N}$ state, consistently with the observed stability of the B centers at long times.

Another possible scenario would involve the recombination of a boron interstitial with the UV center to form a B center: $\mathrm{C_B C_N} + \mathrm{B}_i \rightleftharpoons (2\mathrm{ C}_i)_\mathrm{N} $. We do not retain this scenario since it would require the unlikely presence of interstitials at a high concentration so to allow conversion of UV centers with a high yield. Additionally, since boron interstitials are mobile in hBN\cite{Weston18} and the UV-B conversion is reversible, the long-time stable state upon migration would be the UV center $\mathrm{C_B C_N}$, which is inconsistent with observations. Therefore, we believe that the first scenario describing the dissociation of the in-plane carbon dimer into a vertical dimer and a vacancy, followed by migration of the vacancy, explains both the partial reversibility at short times, and the long-time stability. Furthermore, due to the sizably higher mobility of the boron vacancy as compared with the nitrogen vacancy,\cite{Zobelli07, Weston18} we favor the choice of $(2\mathrm{ C}_i)_\mathrm{N}  + V_\mathrm{B}$ rather than $(2\mathrm{ C}_i)_\mathrm{B}  + V_\mathrm{N}$ as the final product of the electron-beam-induced conversion.

\section*{Conclusion}
We have observed electron-beam induced interconversion between individual UV centers and B centers, appearing as abrupt anticorrelated activation or deactivation events of the point defect emission on simultaneously recorded CL scans. The sensitivity at the single-emitter level, combined with the spatial resolution required to track individual emitters, makes it possible to overcome the limitations inherent to ensemble CL measurements. In ensemble experiments such as those realized in prior work,\cite{Roux22, Nedic24, Bianco25} the CL signal is generally not directly proportional to the actual defect concentration due to irradiation-induced modifications of the excitation and collection efficiencies during the measurement. Such effects can hinder the direct observation of one-to-one interconversion between B and UV centers, as we discuss in the Supporting Information section~S3.

The direct observation of electron-induced switching between two widely studied color centers shed new light onto their long debated microscopic structure and their mutual relation, consolidating the recent hypothesis that B centers are vertical carbon dimers. Our results provide a direct explanation for the well-known fact that B centers can only be activated in hBN crystals exhibiting the 4.1~eV spectral signature of the UV centers, and suggests a possible activation scenario, where the electron beam would modify the dimer orientation, which would then be made stable by evacuating the leftover vacancy.

In addition, the ability to observe individual activation events has allowed us to make use of the CL signal as a marker heralding the successful activation of single B centers. We illustrated this by generating an array of eight sites with a strongly sub-Poissonnian distribution of the SPE numbers. The rare and undesired double creations can be fixed by selective photobleaching, a laser-induced process that is also heralded by a visible change of the emission signal. The combination of these two activation and deactivation monitoring techniques allows to generate arrays with only one emitter per site. Our results deepen the knowledge of optically active deep defects in hBN and provides clear pathways towards the deterministic fabrication of top-down nanophotonic devices for applications to optical quantum information. We expect this discussion to trigger further theoretical and experimental studies that could confirm, refine or modify the proposed conversion scenario.
\\

\section*{Methods}
\subsection*{Sample fabrication} \label{Samplefabrication}
Our sample consists of single crystals of high-pressure, high-temperature grown hBN.\cite{Taniguchi07} This material contains native carbon impurities.\cite{Onodera19, Gale22} The flakes are exfoliated on a SiO$_{2}$(300 nm)/Si substrate. After acetone and isopropanol cleaning (5~min each), the sample is annealed at $1000 \ ^{\circ}$C under 200~sccm nitrogen (N$_{2}$) flow during 1~h and left inside the oven during 3~h for cooling. This technique incentivizes the formation of UV emitters.\cite{Gale22} The sample is then cleaned again in acetone and isopropanol (5~min each) in an ultrasonic bath.

\subsection*{B center generation} \label{B center generation}
The B centers are created by 3~kV irradiation and an electron current of $I = 0.01 -0.1$~nA in a commercial scanning electron microscope in high vacuum conditions ($P < 10^{-4}$ Pa) at room temperature. The CL scans are $64 \times 64$ pixels, sampling a region of $1.6$~$\mu$m $\times 1.9$~$\mu$m. The electron irradiation dose is approximately $1.6 \times 10^{7}$ electrons per pixel at each scan. In the single-emitter creation experiment (Fig.~\ref{Timetrace}), the average dose per emitter created is $7.0 \times 10^{8}$ electrons per emitter. In the single SPE array experiment, (Fig.~\ref{Array}), the average dose per emitter created is  $4.8 \times 10^{8}$ electrons per emitter.

\subsection*{Optical characterization} \label{optical characterization}
Photoluminescence measurements are performed in a room-temperature confocal microscope using a NA = 0.95 air objective. The emitters are excited using a PicoQuant laser diode emitting at $405$~nm, and working in either pulsed or cw regime. The photoluminescence signal is filtered using a 17~nm bandpass fluorescence filter centered at 434~nm, and collected by two avalanche photodiodes (Micro Photon Devices PDM and PDM-UV) in the Hanbury Brown and Twiss configuration.

Intensity autocorrelation is measured for all SPEs to ensure that we only consider individual emitters (\textit{i.e.} $g^{(2)}(0)<0.5$). The $g^{(2)}$ functions are fitted using multi-peak functions of the form
\begin{equation}
g^{(2)} = A \sum_{i \in \mathbb{Z}} c_{i} \exp(-|\tau - T_i|/T_1) + B
\end{equation}
where $T_i = iT_\mathrm{rep}$ with $T_\mathrm{rep}$ the laser period of 13.1~ns, $T_1$ the emitter lifetime, and $c_{i}$ is the amplitude coefficient for the period $i$. $A$ and $B$ are fit parameters accounting for the signal and background intensities. All coefficients $c_{i}$ are fixed to 1 except from the central period coefficient $c_0$, which is the fit parameter providing the value $g^{(2)}(0)$. 

Note that our PL setup does not allow for a characterization of UV centers. However, they have been already studied under non-resonant excitation in prior work.\cite{Vuong16, Plo25}\\

\section*{Associated content}

The present work is associated with a preprint. A. N\'u\~{n}ez Marcos, C. Arnold, J. Barjon, S. Buil, J.-P. Hermier, A. Delteil, Deterministic generation of single B centers in hBN by one-to-one conversion from UV centers, 2025, 2511.02465. arXiv. https://arxiv.org/abs/2511.02465

\section*{Data availability}

The data generated in this study are available at 10.5281/zenodo.19110631.

\section*{Supporting information}

Supporting information available: additional CL maps under different irradiation conditions; additional data about controlled photobleaching; time-resolved ensemble CL measurements; SEM image of a generated array.

\section*{Acknowledgments}
 This work is supported by the French Agence Nationale de la Recherche (ANR) under references ANR-21-CE47-0004 and ANR-25-CE47-3104. This work received government funding under France 2030 (QuanTEdu-France) with reference number ANR-22-CMAS-0001. The Authors acknowledge K. Watanabe and T. Taniguchi for providing hBN crystals, and thank J.-M. Chauveau and C. Vilar for fruitful discussions.

\pagebreak
~
\newpage

\onecolumngrid
\begin{center}
  \textbf{\large Supporting Information\\~\\ Deterministic generation of single B centers in hBN by one-to-one conversion from UV centers}\\[.2cm]
 Andr\'es N\'u\~nez Marcos, Christophe Arnold, Julien Barjon, St\'ephanie Buil, Jean-Pierre Hermier, Aymeric Delteil\\[.1cm]
  {\itshape \small Universit\'e Paris-Saclay, UVSQ, CNRS,  GEMaC, 78000, Versailles, France. \\
{\color{white}--------------------} aymeric.delteil@usvq.fr{\color{white}--------------------} \\}

\end{center}

\setcounter{equation}{0}
\setcounter{figure}{0}
\setcounter{table}{0}
\setcounter{page}{1}
\renewcommand{\theequation}{S\arabic{equation}}
\renewcommand{\thefigure}{S\arabic{figure}}
\section{S1. Additional CL maps}

\begin{figure}[h]
 \hspace*{-0.1\linewidth}
 \includegraphics[width=1.2\linewidth]{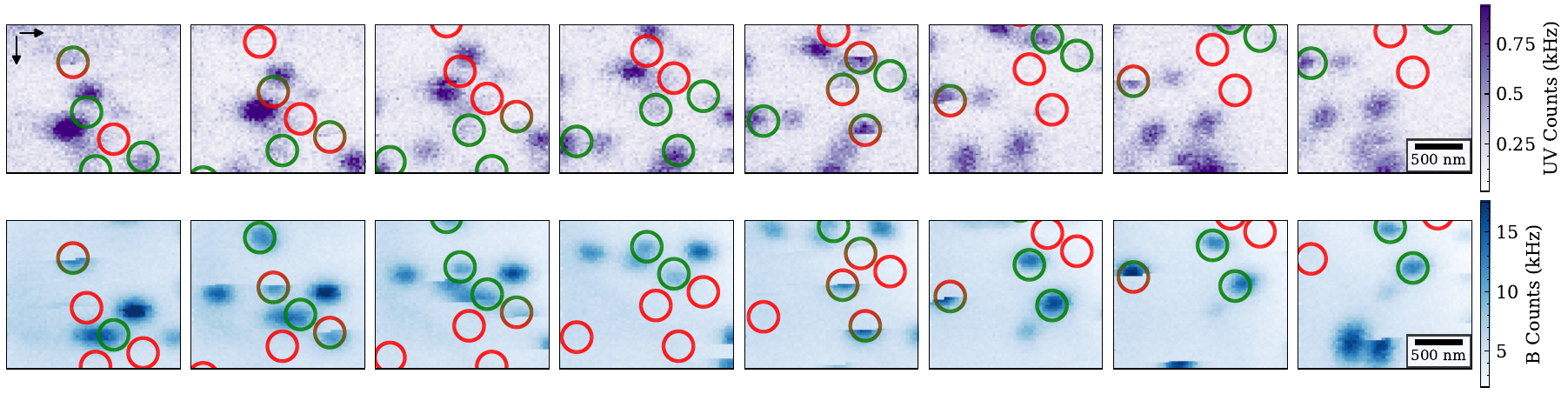}%
 \caption{\label{MoreCLmaps} Consecutive CL mapping of individual UV and blue emitters in a 145~nm hBN flake with a current intensity of $I = 0.024$~nA. The scale bar indicates 500~nm.}%
 \end{figure}

In this section we provide eight additional consecutive CL maps recorded on the 145~nm flake (Figure~\ref{MoreCLmaps}). As in the main text, the circles mark some emitters with the same color code (\textit{i.e.} green for active emitter, red for inactive, two-colored when the state is switching). The measurement protocol is identical to that of Figure~2 of the main text. A minor difference is that the scanned region drifted noticeably during the measurement, which we did not compensate for.

\begin{figure}[h]
 \hspace*{-0.1\linewidth}
 \includegraphics[width=0.7\linewidth]{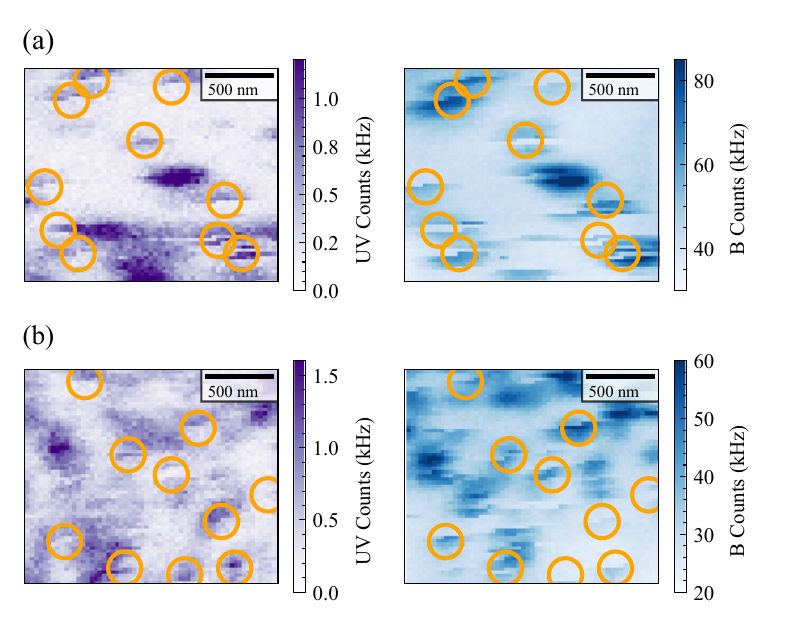}%
 \caption{\label{high_current} Blue and UV simultaneous CL maps at higher current: (a) 0.65~nA and (b) 1.16~nA. The circles indicate some switching events between UV and B centers.}%
 \end{figure}

In addition, on Fig.~\ref{high_current} we show CL maps recorded under a much larger current, up to 1.16~nA. Several switching events between UV and B centers are visible on the same map, with some emitters undergoing multiple switching events during the scan.

\begin{figure}[h]
 \hspace*{-0.1\linewidth}
 \includegraphics[width=0.7\linewidth]{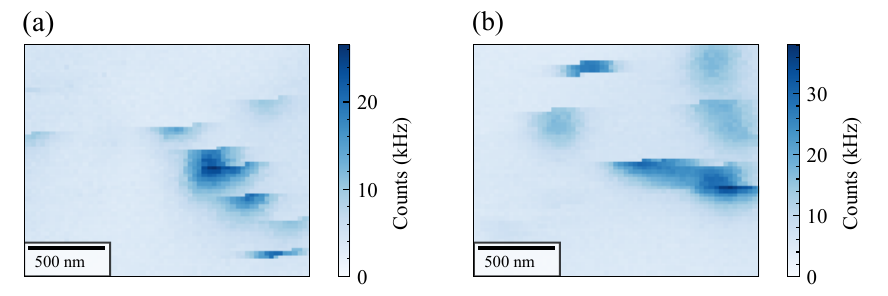}%
 \caption{\label{first_scan} Two CL maps in the blue wavelength range performed on non-irradiated areas, showning B centers appearing during the scan a 145~nm hBN flake with a current intensity of $I = 0.024$~nA. The scale bar indicates 500~nm.}%
 \end{figure}

Finally, we show two blue-range scans of non-irradiated areas, which show that the sample is initially devoid of B centers (Fig.~\ref{first_scan}). They appear during the scans, which translates in spots where the upper part is missing. All CL maps are scanned from left to right, then from top to bottom.

\newpage
\section{S2. Controlled deactivation by photobleaching}

\begin{figure}[h]
 \includegraphics[width=\linewidth]{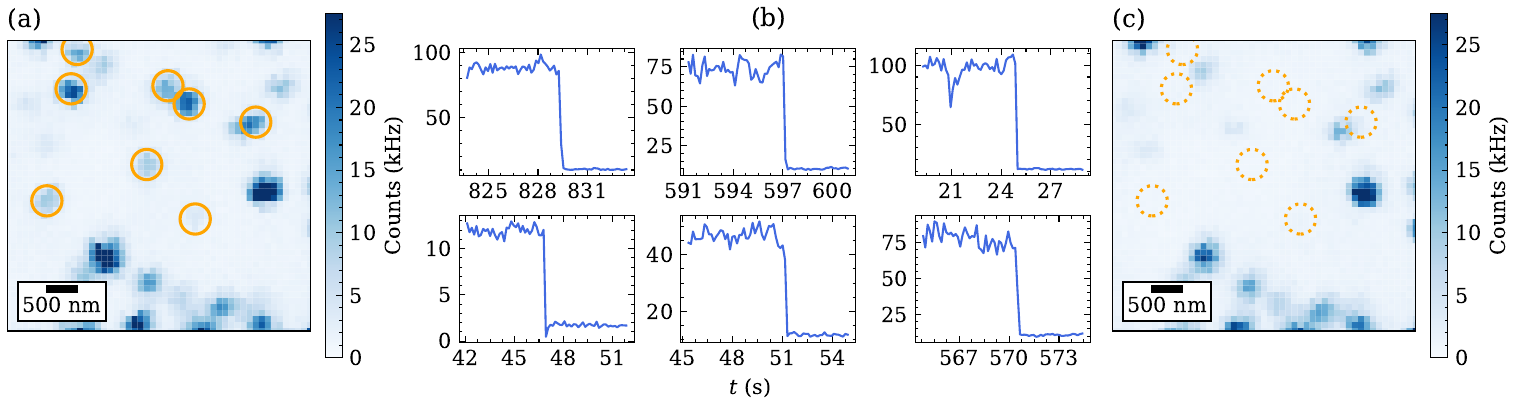}%
 \caption{\label{MorePhotobleaching} 
 (a) PL mapping of random B centers created by continuous electron irradiation in a 90~nm thick hBN flake. (b) PL intensity traces from some of the circled areas in the PL map, showing deactivation of B centers. (c) PL mapping of the same area after selective laser irradiation at high power.}%
 \end{figure}

In this section, we further illustrate the principle of controlled photobleaching. Figure~\ref{MorePhotobleaching}a shows a confocal map of a pre-irradiated area (with randomly positioned emitters). We selected a few emitters and applied the same protocol as in Section III of the main text. Figure~\ref{MorePhotobleaching}b shows six examples of timetraces recorded during the 7.5~mW laser illumination. All show an abrupt bleaching of the signal after a few tens to hundreds of seconds. A confocal map recorded after the selective bleaching is shown on Figure~\ref{MorePhotobleaching}c, where the missing emitters are denoted by the dashed circles.

 \newpage
\section{S3. Individual vs ensemble CL measurements}

In this section, we compare the CL characterization of emitter population dynamics in the regime where individual emitters can be resolved, with the prior approaches based on ensembles~\cite{Roux22, Nedic24, Bianco25}, where the implicit assumption is that the CL signal is proportional to the defect population. We first show that the excitation and detection efficiency is in general not constant, before studying how this affects the inferred population kinetics in ensemble measurements.

\subsection{Time-dependent excitation and collection efficiency}

\begin{figure}[h]
 \includegraphics[width=0.95\linewidth]{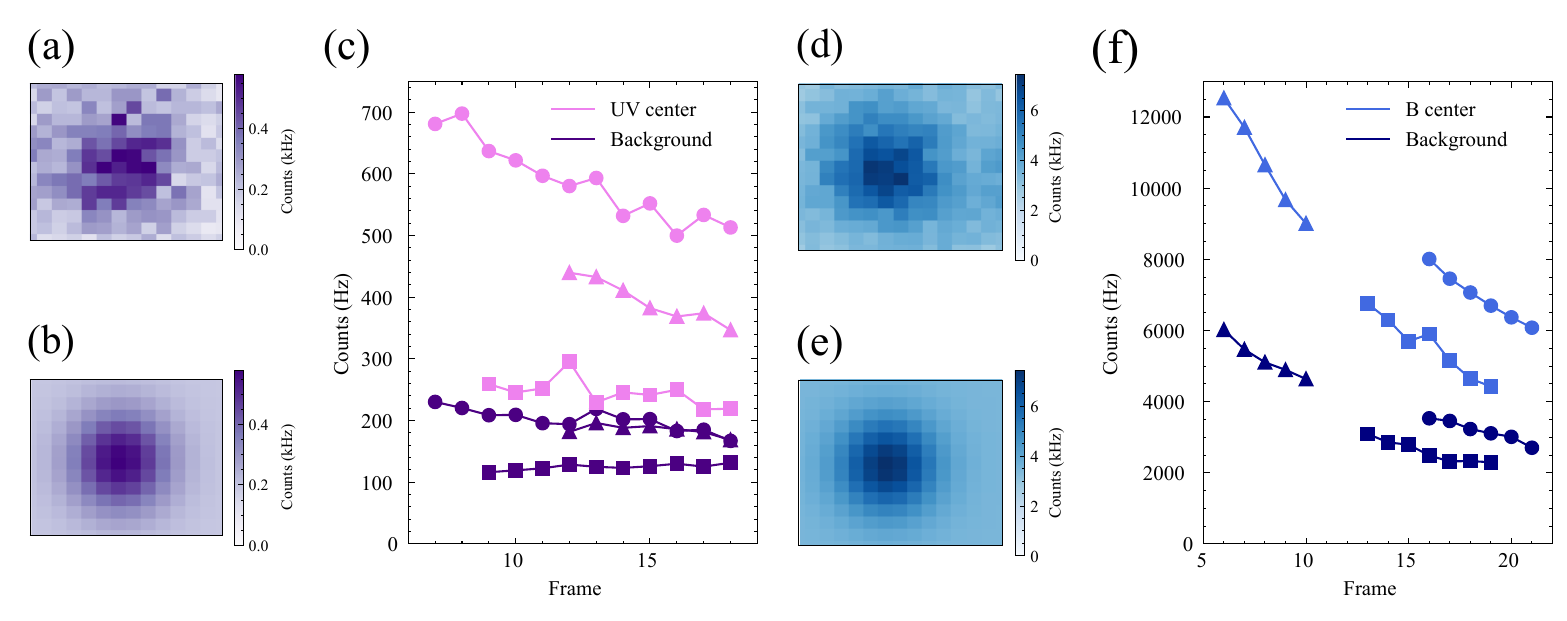}%
 \caption{\label{following_centers} 
 (a) CL map of an individual B center. (b) Gaussian fit to the data. (c) Time evolution of the fit parameters as a function of the frame number for 3 UV centers. Each symbol corresponds to a different color center. (d) CL map of an individual B center. (e) Gaussian fit to the data. (f) Time evolution of the fit parameters as a function of the frame number for 3 B centers. Each symbol corresponds to a different color center.}%
 \end{figure}

The efficiency with which an emitter is excited and its light collected during CL measurements typically evolves in time, due to e.g. irradiation-induced surface contamination by carbonaceous adsorbates (see section~S4), creation of competitive non-radiative recombination channels, and drift of the electron beam position. Fig.~\ref{following_centers} show the amplitude and background associated with single B and UV centers as extracted from Gaussian fit of consecutive CL maps. The intensity of both B and UV centers decrease by 10 to 20~\% over the course of the measurement. A confirmation of this time-dependent efficiency is also visible in Fig.~3 of the main text, where the plateaus associated with the single steps exhibit a slight negative slope.

The main consequence of this observation is that the CL signal associated with a given defect species is not purely proportional to the population, but takes the form $S(t) = \eta(t) N(t) + B(t)$, where $S(t)$ is the CL signal, $\eta(t)$ is the time-dependent efficiency, $N(t)$ the population and $B(t)$ the background. Note that the latter is also time-dependent.

\newpage

\subsection{Monte-Carlo simulations}

\begin{figure}[h]
 \includegraphics[width=0.8\linewidth]{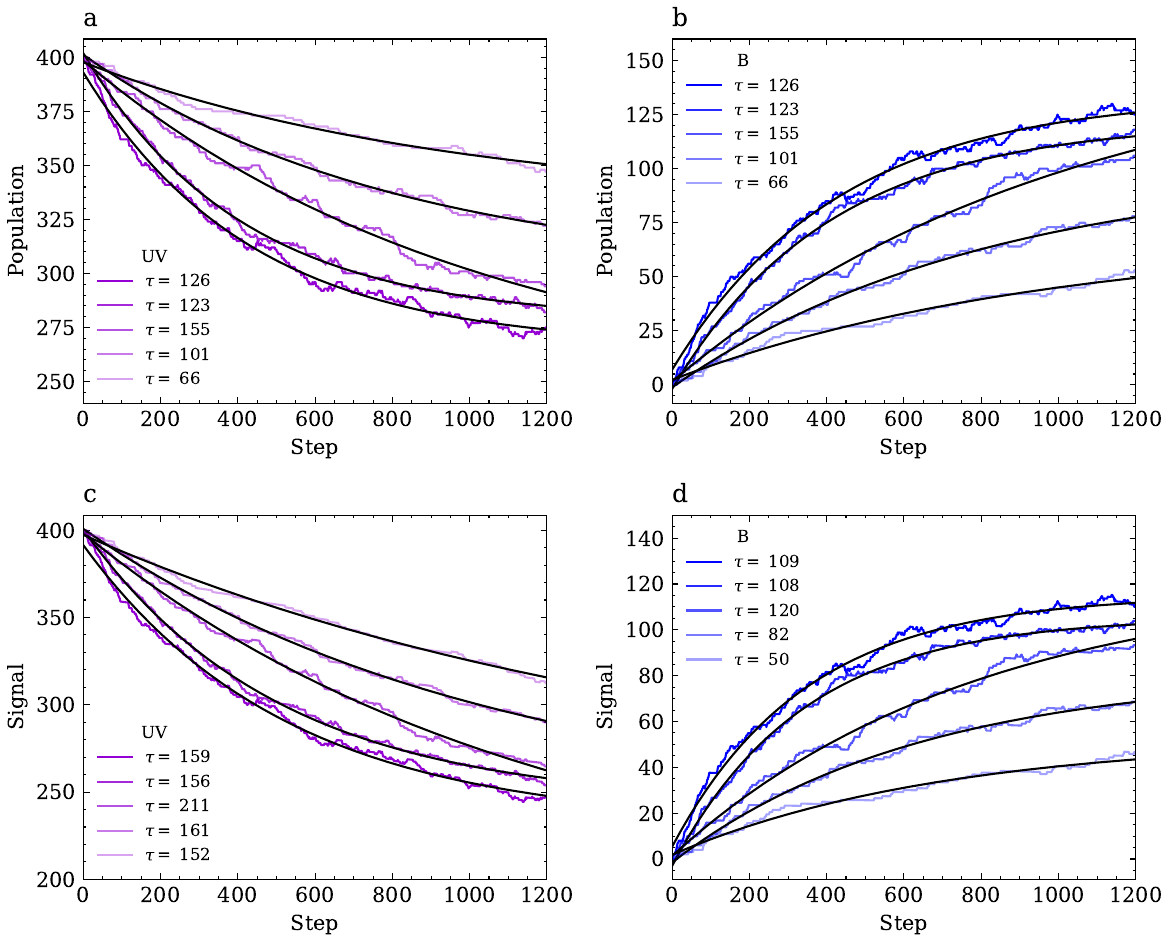}%
 \caption{\label{simu} 
 (a) Simulated populations of UV and (b) B centers at different conversion rates. (c) Simulated CL signal of UV and (d) B centers upon decrease of $\eta$ by 10~\% over the course of the simulation. The black lines are monoexponential fits to the data. The extracted (uniteless) timescales are provided in the appropriate legend. Simulation parameters: UV to B conversion rate from $1.6 \times 10^{-3}$ per step (dark curves) to $0.3 \times 10^{-3}$ per step (light curves). The reverse conversion rate (B to UV) is taken as half the forward rate. The initial population is 400 UV centers and no B center. The simulation is run over 1200 steps.}%
 \end{figure}

In order to qualitatively evaluate the influence of $\eta(t)$ on ensemble measurements, we perform Monte-Carlo simulations of the CL signal. We consider a partially reversible conversion B ~$\rightleftharpoons$~UV. Fig.~\ref{simu}a and b show the simulated populations $N(t)$ as a function of time for three different conversion rates (the simulation parameters are described in the caption). In all cases, we observe time-dependent populations exhibiting anticorrelated fluctuations between the UV and B populations. All curves are well fitted by a monoexponential function, which provides identical time constants for the UV and B population kinetics.

We then plot the associated signal $S(t)$ in the realistic hypothesis that $\eta$ decreases by about 10~\% during the course of the measurement. Fig.~\ref{simu}c and d show the calculation results. The CL signal is only slightly modified by the introduction of $\eta(t)$. However, the main consequence is that the results of a monoexponential fit provides vastly different time constant between the UV and B defects, with the UV decay time being found longer than the B decay time (up to a factor~3). This is due to the variations in $\eta$ increasing (decreasing) the curvature of blue (UV) signal. The exponential fits are very sensitive to these slight changes in curvature and yield very different optimal parameters. In the next subsection, we further illustrate this phenomenon with experimental data.

\newpage

\subsection{Ensemble measurements}

\begin{figure}[h]
 \includegraphics[width=\linewidth]{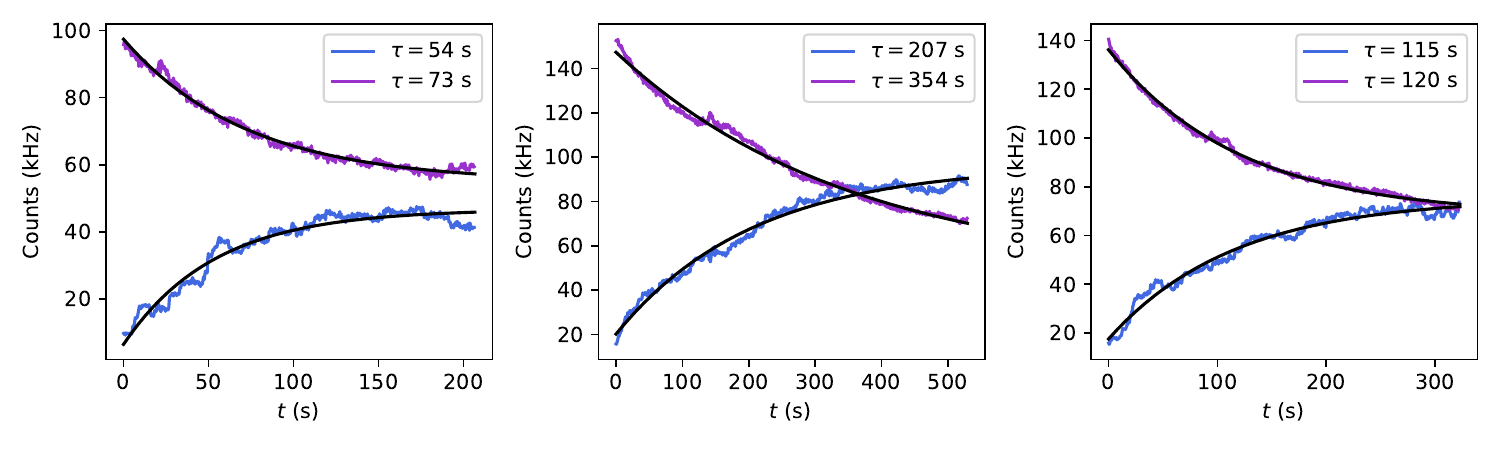}%
 \caption{\label{timetraces} 
 CL time traces of UV (violet curves) and B centers (blue curves) at three different locations. The black lines are monoexponential fits to the data. The extracted timescales are provided in the appropriate legend. }%
 \end{figure}

We perform ensemble CL measurements on a crystal possessing a high density of UV centers. We use a focused beam of current 10 to 20~pA during 200 to 500~s. The results are shown on Fig.~\ref{timetraces}. As in the simulations, the UV and B signals exhibit opposite time evolutions, with the blue signal increasing while the UV signal decreases. Clear anticorrelations are also visible in the signal fluctuations, with many occurrences of discrete jumps in opposite directions in B and UV signals. However, when fitting the signal with a monoexponential function, different time constant are found for the blue and UV signal, with the timescale associated with the UV signal being consistently longer than that of the blue signal, in agreement with the simulations. This shows a limitation of ensemble measurements when evaluating the kinetics of luminescent species in CL. Our approach, based on the identification of individual emitters using enhanced detection sensitivity associated with spatial resolution, allows access to individual particles, which can then be followed in time, and therefore makes it possible to circumvent the limitations of ensemble measurements.

\newpage

\section{S4. Surface contamination during irradiation}

\begin{figure}[h]
 \includegraphics[width=0.3\linewidth]{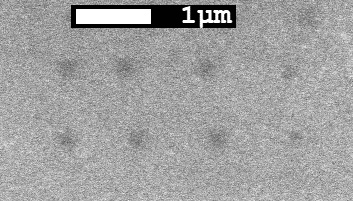}%
 \caption{\label{carbon} post-irradiation SEM image of a $4 \times 2$ array of single B centers. The dark spots at the irradiation sites are attributed to carbon deposition.}%
 \end{figure}

The irradiation process is well known to generate adsorption of carbon-based contaminants on the sample surface. While in-situ monitoring allows to minimize it by stopping the irradiation immediately after the emitter is created, we nevertheless observe a slight contamination at the position of the irradiation sites (Fig.~\ref{carbon}).

\end{document}